\documentstyle[a4,12pt,epsf]{article}

\advance \textheight by \topskip
\begin{document}

\title{Events registration in a fast neutrons spectrometer}

\author{J.~N.~Abdurashitov, V.~N.~Gavrin, A.~V.~Kalikhov,
{\underline{A.~A.~Shikhin}}, \\ V.~E.~Yants and O.~S.~Zaborsky}

\date{\it{Institute for Nuclear Research, Russian Academy of Sciences, Moscow, Russia}}

\maketitle

\begin{abstract}

The principle of operation, design, registration system and main
characteristics of a fast neutrons spectrometer are described. The
spectrometer is intended for direct measurements of ultra low fluxes of fast
neutrons ~\cite{abd1}. It is sensitive to neutron fluxes of
10$^{-7}$ cm$^{-2}$s$^{-1}$  and  lower. The detection efficiency of fast neutrons with
simultaneous energy measurement is within 0.03--0.09 for neutron energies $>$0.7 MeV
and depends on the neutron energy and the spectrometer response function.

The neutron spectrometer was designed taking into account requirements for
minimizing the $\gamma$-ray and random coincidence backgrounds. It is a
calorimeter based on a liquid organic scintillator-thermalizer with helium
proportional counters of thermalized neutrons distributed uniformly over the
volume. The energy of thermalized neutrons is transformed into light signals
in a scintillation detector. The signals from proportional counters provide
a ``neutron label'' of an event.

Low-level signal electronics for the spectrometer were designed with
signal-to-noise ratio optimization and full pulse shape analysis
required for efficient rejection of background events. A data
acquisition and processing system is based on fast (100 MHz) two-channel
PC/AT interfaced   digital   oscilloscope.   The   acquisition
software was written in the C programming language.

\end{abstract}

\section{Principle of operation  and design of the spectrometer}

\begin{figure}
\epsfsize=0.5\textwidth
\epsffile{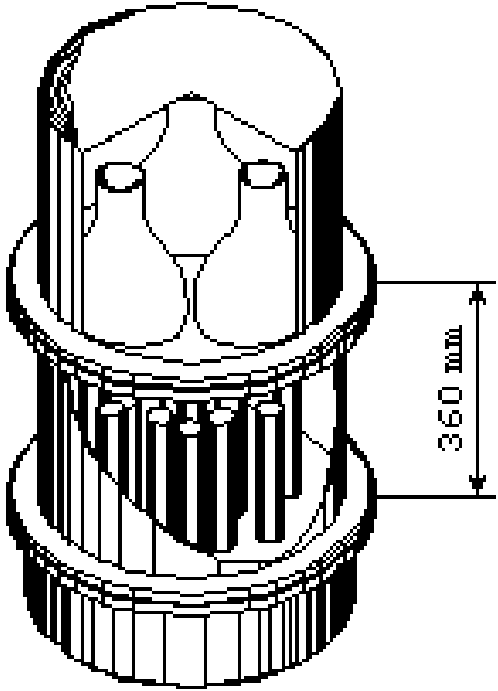}
\caption{General view of the detector}
\label{f1}
\end{figure}

The detection part of the spectrometer (detector) consists of an organic
scintillator viewed by photomultipliers (PMT) and proportional counters with
$^3$He (neutron counters --- NC) distributed uniformly over the
scintillator volume. Figure ~\ref{f1} shows a general view of the detector. Fast
neutrons (E$_n >$1 MeV) enter the scintillator, are decelerated down
to thermal energy, and diffuse in the detector volume until they are
either captured in a neutron counter or captured by scintillator protons or
leave the detector. The amplitude of light scintillations from recoil
protons, which are produced during neutron thermalization, is on average
proportional to the initial neutron energy, if the energy losses due to
scattering by carbon and a non-linear dependence of the scintillator light
yield on the energy of recoil protons are neglected. A portion of
thermalized neutrons in the neutron counters is captured by $^3$He
nuclei, which emit protons and tritium nuclei.

To simplify the apparatus structure, signals from all PMTs and counters are
multiplexed into independent channels called ``PMT channel'' and ``NC
channel'' respectively. A signal from NC channel triggers the data
acquisition system. The full waveforms of events in the PMT and NC channels
are registered independently inside selected time interval before and after
trigger, which are called ``prehistory'' and ``history'' accordingly, by
means of two-channel digital oscilloscope. This time interval can be
adjusted in the acquisition algorithm over the wide range from 0.2 $\mu$s
to 2.6 ms and is selected on the basis of calibration measurements.

There is a certain distribution  of  time  intervals  between appearances
of a signal from PMTs and a signal from one of the NCs related
with one neutron, which is associated with
specific characteristic feature of the detector --- delay. This delay is
conditioned by the mean lifetime of thermalized neutrons inside the detector
volume and is determined mainly by the detector design. If at least one
event occurs in the PMT channel during the acquisition time interval, it is
``labelled'' as the signal coinciding with neutron capture in the NC. The
amplitude of the labelled PMT signal is considered as a measure of the initial
neutron energy. This technique for collecting events allows one to
significantly (by several orders of magnitude) suppress the natural
background of $\gamma$-rays.

The detector housing is made of stainless steel. The volume of cylinder tank
for scintillator is about 30 l (taking into account the volume of NCs). To
enhance light   collection,   the  walls  and  bottom  of  the
scintillator tank were coated with a  fluoroplastic,  and  19 wells for NCs were coated with aluminium
foil. An illuminator made of organic glass was covered by fluoroplastic disk
with windows  intended  for  improving  light  collection  and securing three PMTs
Model 173.  PMTs,  preamplifiers,  and  a  summator  of  PMTs signals were placed
under the light-protecting cover.

CHM-18 neutron counters filled with a ($^3$He + 4\%Ar) mixture at
a pressure of 4 atm are mounted in sealed wells in the interior volume of
the tank filled with the scintillator. The anode contacts of the counters
are terminated under the lower tank plane. The preamplifiers of signals from
NCs, and the signal summator are placed in the bottom part of the detector
and are also covered with a cap. The Protva NE-213 scintillator used in the
detector has the following characteristics: 0.84 g/cm$^3$ (density); 136
g/kg (hydrogen concentration); 3$\pm$0.6 ns (scintillation duration); a
light yield of at least 40\% of anthracene; and 80$^\circ$C (combustion temperature).

\section{Low-level signal electronics}

\subsection{Formation of the Initial Signals}

Signals from the PMTs and NCs are amplified by the preamplifiers (PA),
are multiplexed in the linear summators, and are entered into the data acquisition
system. The circuit design of the low-level signal electronics are selected
taking into account an optimal signal-to-noise ratio. Their basic
characteristics are listed in the Table~\ref{t1}.

\begin{table}
\caption{Basic performance data of the preamplifier and linear summators}\label{t1}
\begin{center}
\begin{tabular}{|l|c|c|}
\hline
Characteristics & Preamplifier & Summator NC (PMT) \\ \hline
Gain (K) & 2.5--5, variable & inverting $-$1 ($-$1--5, variable) \\ \hline
Max. out voltage, V & $-$2 & +2 (+3.5) \\ \hline
Load impedance, $\Omega$ & 50 & 50 \\ \hline
Risetime, ns & 16 (K=5), 6 (K=2.5) & 9 (3) \\ \hline
Bandwidth &  100  Hz  --  20  MHz  (K=5)  & 100 Hz -- 40 (100) MHz\\
\cline{2-2}
& 100 Hz -- 55 MHz (K=2.5) &  \\ \hline
Output noise (PA  on) &  & $\sim$3 ($\le$1) mV, 2U$_{max}$ \\
\cline{1-1}
\cline{3-3}
Dynamic range &  & $\ge$500 (3500) \\ \hline
Power requirements & $\pm$12 V, $\pm$16 mA & $\pm$12 V,+111/$-$105(+25.5/$-$22) mA \\ \hline
\end{tabular}
\end{center}
\end{table}

The anode circuit's capacitance of the PMT Model 173 is $\sim$20 pF, and the time
of electron collection by the anode is $\sim$90 ns. The PMT anode load is 1
M$\Omega$. Since the PA bandwidth ensures a sufficient margin with respect
to the initial signal spectrum, an exponential pulse with a duration $\sim$80 $\mu$s
and a trailing edge time constant of $\sim$20 $\mu$s
forms at its output.

\begin{figure}
\epsfsize=0.8\textwidth
\epsffile{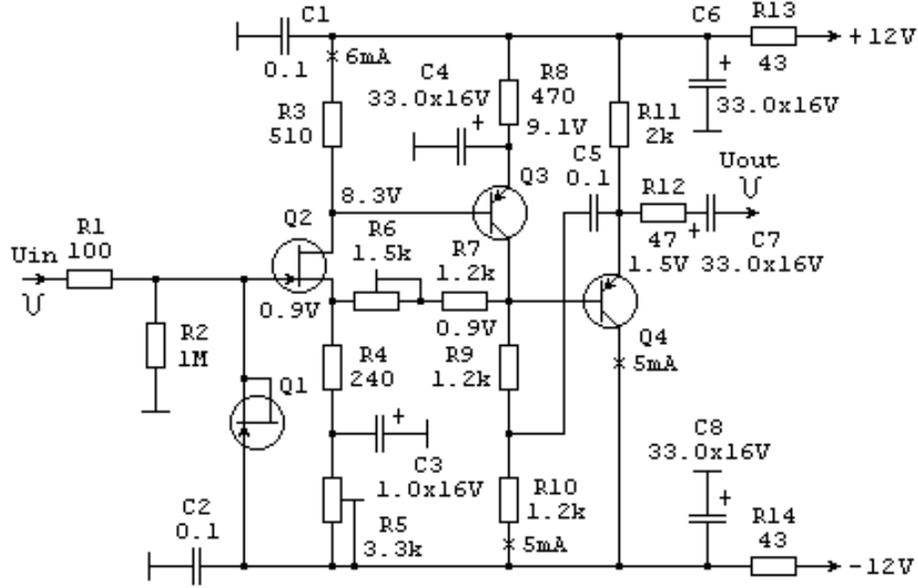}
\caption{The circuit diagram of a fast low-noise preamplifier}
\label{f2}
\end{figure}

Difficult problem is to provide the necessary signal-to-noise ratio in the
NC channel. Joining signals from 19 counters into a common channel
increases the noise at the linear summator output by a factor of at least
4.5. To reduce the noise, the main amplification of initial signals is
accomplished with the preamplifiers before their joining. Figure ~\ref{f2} shows the
circuit diagram of a fast low-noise preamplifier, which was developed
especially for this purpose.  It is used in both channels  of the spectrometer:
PMT and NC.

\subsection{Preamplifier}

The preamplifier has a two-stage common-source-common-emitter circuit with
common serial negative-voltage feedback loop (NFL) and directly coupled
stages. Each stage has a local serial, current NF. The first stage is based
on a  FET  with normalized coefficient of the noise charge and
has gain of 2,  which also improves the noise characteristics
of the PA as a whole. The third stage of PA is an emitter follower with bootstrapping over C$_5$
to common point of R$_9$ and R$_{10}$, which
increase PA gain without common NFL 4--5 times. The total gain of PA with an
open NFL is $\sim$3750. For a closed NFL, the gain is continuously
adjusted from 5 to 10, and the minimum amount of feedback is $\sim$50
dB. The parameters of PA circuit elements are selected so as to minimize the
potential difference between the Q$_2$ source and Q$_3$
collector and thus to reduce the effect of parasitic capacitance of NFL
elements on the PA speed. Its dc operating mode and gain are set with a
trimmer resistors R$_5$ and R$_6$, respectively. A
protective circuit at the PA input is based on a FET Q$_1$
connected as a diode. The input impedance is determined by the R$_2$
value and equals 1 M$\Omega$ in our case.

An output emitter follower isolates the NFL and load circuits and transmits
pulses from PA to summator through an RG-174 coaxial cable with a 50 $\Omega$
characteristic impedance. Power matching conditions used in the device
decrease the final gain by a factor of 2 without an appreciable decrease in
the signal-to-noise ratio, but protect the PA output stage from shorting in
the load, improve the linearity, provide the minimal distortions of the
signal waveform, and simplify the spectrometer assembly, allowing for
optimal wiring of circuits.

\subsection{Linear Summators}

\subsubsection{NC channel}

\begin{figure}
\epsfsize=0.8\textwidth
\epsffile{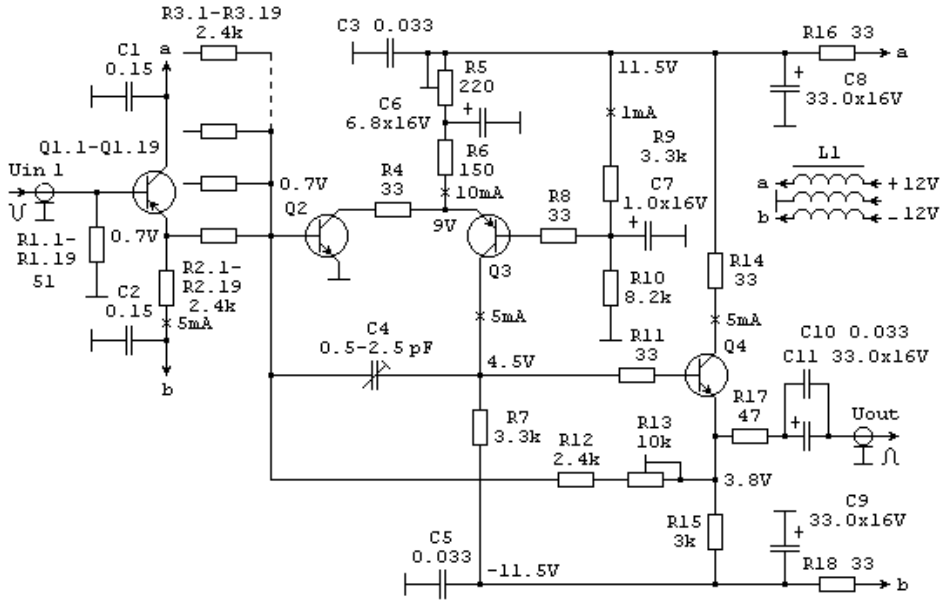}
\caption{The circuit diagram of a fast low-noise linear summator (NC)}
\label{f3}
\end{figure}

To attain the maximum signal-to-noise ratio, signals from NC preamplifiers
were joined simultaneously in one stage for all 19 channels by using a fast
low-noise linear summator which electric circuit diagram is shown in
Figure ~\ref{f3}.

The summator has a complementary cascode circuit with an output emitter
follower on transistors Q$_2$--Q$_4$ with a common
parallel voltage NFL. The summator operation is based on the known principle
of adding the input current signals, which are specified by the weight
resistors R$_{3.1}$--R$_{3.19}$, at a small input impedance of
the amplifier with a parallel NFL applied.

The input impedance of the cascode circuit is 250 $\Omega$ for a collector
current of 5 mA and a typical h$_{21e}$=50. The amplifier gain with
an open NFL is $\sim$660. The gain with a closed NFL is set at 2 for
each input by using trimmer resistor R$_{13}$. The amount of the
feedback changes as a function of the number of weight resistors R$_3$
from 330 to 17, thus resulting in a variation of the amplifier input
impedance from 0.8 to 15 $\Omega$, i.e., by a factor of almost 20. This
leads to a dependence of the summator gain on the number of operating
inputs, which is extremely unacceptable, because during spectrometer
operation (for example, when performing calibration measurements), PA may be
switched off by switching off supply voltages in groups or individually.

To eliminate this dependence, the emitter followers on Q$_{1.1}$--Q$_{1.19}$
are included in the summator circuit. They isolate the weight
resistors incorporated in the feedback from the input circuits. The summator
input impedance determined by the R$_1$ resistance value is 50 $\Omega$.

The potentials at emitters of input followers and at the Q$_2$ base
are almost equal. Therefore, the voltage drop at the weight resistors is
small, allowing for their dc coupling. Nevertheless, due to a large number
of summed channels, an additional direct current of $\sim$0.5 mA appears in
the feedback and produces a voltage drop of $\sim$3 V across resistor R$_{12}$
and R$_{13}$. This factor was taken into consideration
when calculating the circuit operation for the dc. The trimmer resistor R$_5$
sets the Q$_2$ and Q$_3$ collector currents
equal to each other.

The emitter follower Q$_4$ included in the common NFL creates an
additional pole in the frequency response of the amplifier, decreasing its
upper boundary frequency and increasing the phase shift at higher
frequencies. The frequency response is corrected with a C$_4$
by proceeding an    optimal   form   of   the   transient characteristics.

Pulses with an exponential fall time duration of $\sim$80 $\mu$s are formed at
the output of the linear summator of the NC channel. These pulses have front
duration determined by the features of an event in a counter and range from
1 to 6 $\mu$s.

\subsubsection{PMT channel}

The signals from PMT preamplifiers are joined just as in the NC channel with
only one difference. PA are placed immediately on the PMTs HV
dividers for minimization of parasitic capacitance in their input circuits
and were closed by the light-protecting cover. Therefore, the trimmer
resistors which are placed on the PA printed boards inaccessible for
operative adjustments, and they are used for preliminary adjustments only.
Exact gain equalization of separate PMTs is performed with the help of
trimmer weight  resistors of linear summator (which is mounted under the top of
the light-protecting cover), which axes are available through the top of
the cover under slot.

\begin{figure}
\epsfsize=0.8\textwidth
\epsffile{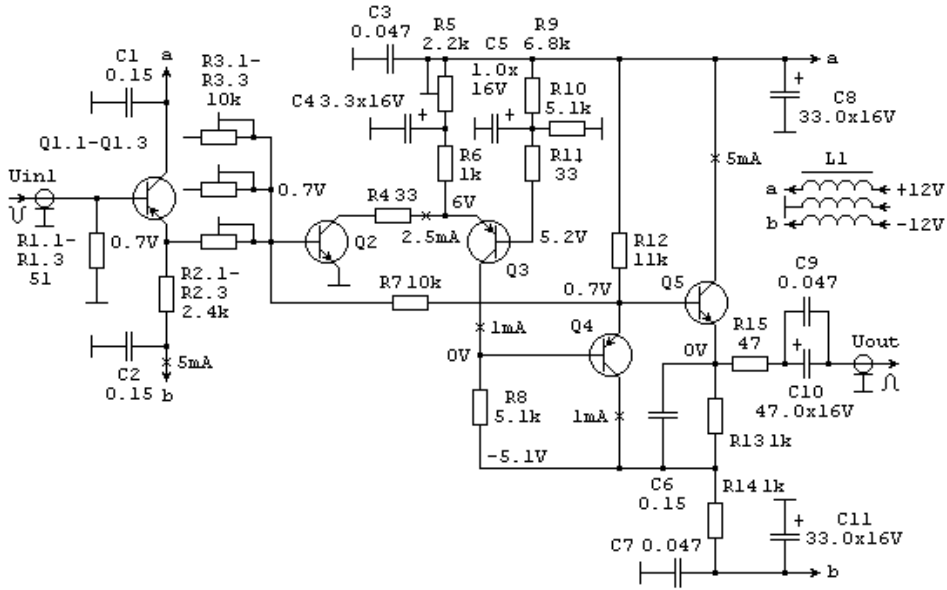}
\caption{The circuit diagram of a fast low-noise linear summator (PMT)}
\label{f4}
\end{figure}

An electric circuit diagram of the fast low-noise linear summator is shown
in Figure ~\ref{f4}. It was made on the base of well-known scheme of Radeka ~\cite{rad1,rad2}
which was optimized for power voltages $\pm$12 V. This is inverting
amplifier with common parallel voltage NFL. It is built on the base of
cascode circuit Q$_2$--Q$_3$ and output stage with
bootstrap on Q$_4$ and Q$_5$.

This choice is conditioned by requirements of more high speed of response
and linearity of the PMT channel, which provides the future possibility of $\gamma$-ray
background discrimination in wide dynamic range with using of current
pulse waveform from PMT anode.

The first stage of the cascode is performed on bipolar transistor Q$_2$
which transconductance S$\simeq$100 mA/V for collector current I$_c$=2.5 mA
(this is optimal mean which is adjusted by trimmer R$_5$), what
exceeds the mean of this parameter for best FETs with factor 5--6. This
approach permitted an easy way to compensate the loss of gain at the expense
of low power voltages of scheme. The gain of amplifier with open NFL is $\sim$12000.
The range of gain adjustment with closed NFL which is
controlled by trimmer weight resistors R$_{3.1}$--R$_{3.3}$ is
equal 1--5 in this case. The minimum amount of feedback is $>$60 dB. The
basic parameters of PMT summator are given in the Table I for the gain equal 3.

\section{Data acquisition system}

\subsection{Overview}

The third generation of the data acquisition system is described here. The first
system was based on slow electronics, a hardware delay line of 80 $\mu$s,
and a multichannel analyzer used for data storage and indication. The second
acquisition system was previously described in detail ~\cite{abd1}.

\begin{figure}
\epsfsize=0.95\textwidth
\epsffile{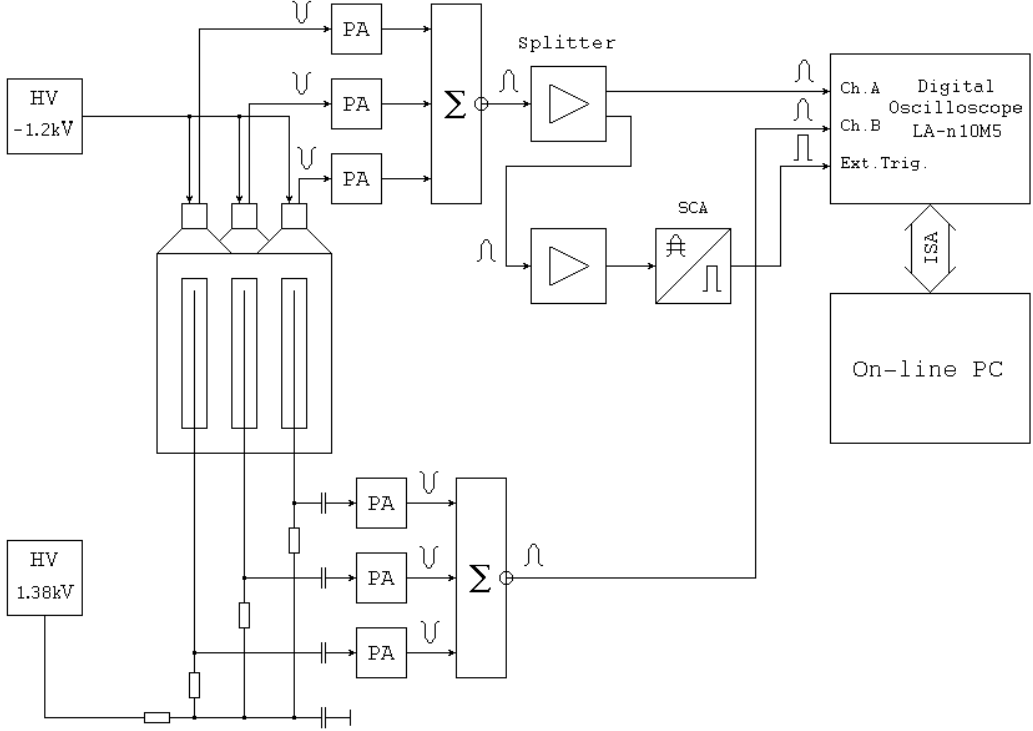}
\caption{The functional diagram of a data acquisition system}
\label{f5}
\end{figure}

A data acquisition system, whose functional diagram is shown in Figure ~\ref{f5},
can be  conventionally  divided  into  three  parts:  the  PMT channel,
the NC channel and measuring part which includes fast (100 MHz) two-channel PC/AT
interfaced digital oscilloscope (DO), and several supplementary units.

Negative signals from the anodes of the three PMTs are entered to the inputs
of the  preamplifiers. The sufficiently  high  speed  of
response and a small input capacitance make it possible to study the feasibility of
$\gamma$-ray  background events discrimination by  their waveforms. A
continuously adjusted PA gain ensures the operation of all three PMTs from a
single high-voltage power supply.

The PMT signals from PA outputs are multiplexed by a fast inverting linear
summator and then ramified on two directions. One branch entered directly
to the input of the first channel of DO. In the other branch, the
PMT signals are entered to the input of combined unit of amplifier and
single-channel analyzer (SCA). The positive TTL-specified signal from SCA
output triggers the DO through an external trigger input.

In the NC channel, a high voltage of positive polarity from a single source
is applied through high-voltage isolating resistors to the anodes of helium
counters. The latter generate signals of negative polarity, which are fed
through high-voltage separating capacitors to the inputs of PAs. The spread
of the counters gas amplification factor is cancelled by adjusting the PA
gain during calibration. For convenient channel tuning, each of them
(similarly to the PMT channel) can be switched off independently by turning
off the PA supply voltages.

The signals of neutron counters from the PA outputs are multiplexed in a
single channel by using a fast inverting linear summator and subsequently
enter directly to the input of the second channel of DO.

The waveform of events in the NC channel is recorded also and ensures the
feasibility of discrimination $\alpha$-particles background by means of
mathematical analysis methods.

\subsection{Digital Oscilloscope}

Two-channel digital oscilloscope model LA-n10M5 (Rudnev-Shil'aev Co., Russia) ~\cite{rud1}
is used as the base of spectrometer data acquisition
system. Its main performance data are listed in the Table~\ref{t2}.

\begin{table}
\caption{Main performance data of the digital oscilloscope model LA-n10M5}\label{t2}
\begin{center}
\begin{tabular}{|l|c|}
\hline
Number of independent A/D channels & 2 \\ \hline
Word length (resolution) of A/D conversion & 8 bit (256 digitalizing levels) \\ \hline
Sample frequency range $^*$ & 3.052 kHz -- 50 MHz (two-channel mode) \\
&--100 MHz$_{max}$ (one-channel mode) \\ \hline
Digitalizing period range $^*$ & 20 ns (two-channel mode) --\\
&10 ns (one-channel mode) -- 0.3277 ms \\ \hline
Volume of internal RAM $^*$ & up to 256 kB \\
&(up to 128 kB/channel) \\ \hline
Input sensitivity $^*$ & $\pm$1V, $\pm$0.5V, $\pm$0.2V or $\pm$0.1V \\ \hline
Input impedance & 1 M$\Omega$ and 15 pF \\ \hline
Signal-to-noise ratio & 47 dB \\ \hline
Coefficient of harmonic distortions & $-$51 dB \\ \hline
\end{tabular}
\end{center}
$^*$ --- programmable value
\end{table}

This unit is a 3/2-sized (103$\times$245 mm$^2$) standard printed board
placed in an arbitrary ISA-bus slot of the PC/AT-compatible computer.

The start of conversion is produced by one of the input analog signals or an
external trigger signal. The synchronization can be performed by edge or
level.

An order of internal cyclic RAM operation is following. After
the start of conversion command the data from an ADC is continuously written to the selected part of
RAM which is called ``prehistory''.
Synchronization pulses are blocked until the volume of prehistory is not filled up.
After prehistory filling and
triggering by the synchronization pulse the part of RAM is written with the
deduction of prehistory volume. This part of RAM is called ``history''.
There is a possibility of switching sample frequency from current value to
50 or 6.25 MHz after finishing prehistory records and arrival
of synchronization pulse.

The data from oscilloscope RAM can be transmitted into computer memory in
the DMA mode.

The acquisition software is written on C language using Borland C$^{++}$
compiler. It use all performance of the unit and operates under DOS
command prompt mode in MS Windows-95 operation system.

\subsection{Operation Algorithm}

As was mentioned above, a certain time delay of the signal from NC relative
to its PMT signal is a characteristic feature of the detector. Therefore, it
is important to select correctly the watching time interval (WTI) for
acquisition system. The delay time distribution is of an exponential type $e^{-t/T}$,
where $t$ is the delay time and $T$ is the mean lifetime of
thermalized neutron in the detector volume. The direct measurement of the
delay time distribution of true neutron events was performed ~\cite{abd1} with a Pu-Be
source. This distribution corresponds to the mean life time value $T$ $\sim$80 $\mu$s.
The WTI should be selected on the basis of this value. Its
particular value depends on background conditions mainly under which
measurements of the neutron fluxes are performed.

There are two basic operation modes of the acquisition software: Pulse
Acquisition Mode and Spectrum Acquisition Mode.

\begin{figure}
\epsfsize=0.8\textwidth
\epsffile{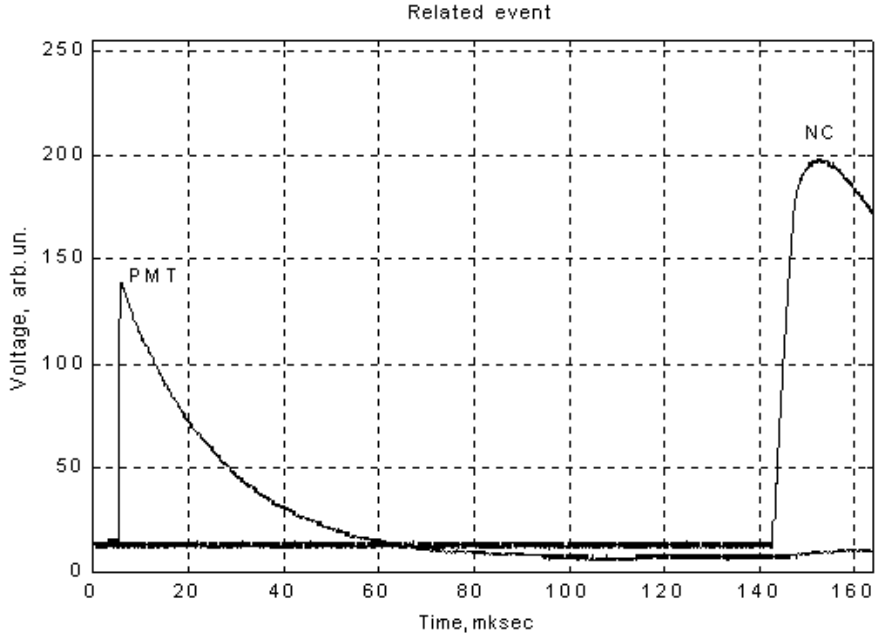}
\caption{The typical related neutron event}
\label{f6}
\end{figure}

Pulse Acquisition Mode is usually used for real background measurements. In this mode
acquisition is triggered by signal from NC channel. Value of WTI equal 164 $\mu$s
is selected what corresponds to 8 kB/channel of the DO RAM for
digitizing period equal 20 ns. Therefore, one event is occupied about 16 kB
of memory.  The prehistory volume is programmed on  the  value 14/16 parts of
WTI, i.e. it is occupied 7 kB of memory or 143 $\mu$s of the time scale.
The full waveforms of events in both channels of DO inside WTI are written
to the hard disk of computer with using a binary format of data. The maximal
counting rate in this case is about 8 events/sec for on-line computer with
133 MHz Pentium-S processor. There is a possibility to decrease the space of
memory which occupied by an event at the expense of information about front
pulse shape of an event in PMT channel. In this case not all information
about an event is written. In the NC channel the frame only with dimensions $\pm$512
bytes from the prehistory/history boundary, which includes front
pulse shape, amplitude and the part of slope of an event is written to the
hard disk  with the digitizing period equal 20 ns.  In the PMT
channel each first, eighth, sixteenth etc. digitizing points
only inside WTI are written. Therefore, this
technique allows one to obtain decreasing factor equal 8. Figure ~\ref{f6} illustrates a
typical ``picture'' of the related neutron event, registered by DO.

Secondly, Spectrum Acquisition Mode, is used for calibration procedures,
system operation stability control and background rate monitoring during
acquisition. In this mode system is externally triggered by signals from
SCA. This performance allows to select the threshold of acquisition more
exactly then in case with using of internal triggering. One-channel mode
with maximal sample frequency equal 100 MHz and 1 kB of DO RAM, what
corresponds to 10 $\mu$s of the time scale, are used. The prehistory volume
is programmed on the value 4/16 parts of the time scale. The amplitude and
histogram (projection onto amplitude scale) spectra of events are
calculated on-line and are written to the hard disk of computer with using an
ASCII format of data. The maximal counting rate in this case is about 120
events/s. During background measurements this mode starts automatically on 5
min. after ``write-to-disk'' command and repeats during this time after each 55
min. of running in the Pulse Acquisition Mode.

Any alteration of DO set-up (contents of INI file), switching between
acquisition modes or start of acquisition with ``write-to-disk'' command is
accompanied by automatic on-line calculation of ``base lines'' real
position. For this purpose DO is switched on short time to automatic mode of
horizontal sweep and minimal value of sample frequency.

\section{Acknowledgments}
We thank G.T. Zatsepin for stimulating interest to the work and useful
discussions, also J.S. Nico and S.V. Girin for careful reading
of this article and their critical remarks. We acknowledge the support of the Russian Foundation of Basic
Research also. This research was made possible in part by grant of RFBR No.~98--02 16962.

\end{document}